\documentclass[iop,apjl]{emulateapj}

\usepackage{amsmath}        
\usepackage{amssymb}

\newcommand{\perbeam}{\,beam$^{-1}$}

\newcommand{\chandra}{{\it Chandra}}
\newcommand{\xmm}{{\it XMM-Newton}}

\shorttitle{No radio emission from an IMBH in G1}
\shortauthors{Miller-Jones et al.}

\begin{document}

\title{The absence of radio emission from the globular cluster G1}

\author{J.~C.~A. Miller-Jones}
\affil{International Centre for Radio Astronomy Research - Curtin University, GPO Box U1987, Perth, WA 6845, Australia}
\email{james.miller-jones@curtin.edu.au}

\author{J.~M. Wrobel}
\affil{NRAO Domenici Science Operations Center, 1003 Lopezville Road,
  Socorro, NM 87801}

\author{G.~R. Sivakoff, C.~O. Heinke and R.~E. Miller}
\affil{Department of Physics, University of Alberta, Room 238 CEB, Edmonton, AB T6G 2G7, Canada}

\author{R.~M. Plotkin}
\affil{Astronomical Institute `Anton Pannekoek', University of Amsterdam, Science Park 904, 1098 XH, Amsterdam, the Netherlands}

\author{R. Di\thinspace Stefano}
\affil{Harvard-Smithsonian Center for Astrophysics, 60 Garden Street, Cambridge, MA 02138}

\author{J.~E. Greene}
\affil{Department of Astronomy, The University of Texas at Austin, 1 University Station C1400, Austin, TX 71712}

\author{L.~C. Ho}
\affil{The Observatories of the Carnegie Institution for Science, 813 Santa Barbara Street, Pasadena, CA 91101}

\author{T.~D. Joseph}
\affil{School of Physics and Astronomy, University of Southampton, Highfield SO17 IBJ, England}

\author{A.~K.~H. Kong}
\affil{Institute of Astronomy and Department of Physics, National Tsing Hua University, Hsinchu 30013, Taiwan}

\author{T.~J. Maccarone}
\affil{School of Physics and Astronomy, University of Southampton, Highfield SO17 IBJ, England}

\begin{abstract}
The detections of both X-ray and radio emission from the cluster G1 in M31 have provided strong support for existing dynamical evidence for an intermediate mass black hole (IMBH) of mass $1.8\pm0.5\times10^4M_{\odot}$ at the cluster center.  However, given the relatively low significance and astrometric accuracy of the radio detection, and the non-simultaneity of the X-ray and radio measurements, this identification required further confirmation.  Here we present deep, high angular resolution, strictly simultaneous X-ray and radio observations of G1.  While the X-ray emission ($L_X=1.74^{+0.53}_{-0.44} \times 10^{36} \, (d/750 {\rm \, kpc})^2  {\rm \, erg \, s}^{-1}$ in the 0.5--10 keV band) remained fully consistent with previous observations, we detected no radio emission from the cluster center down to a $3\sigma$ upper limit of 4.7\,$\mu$Jy\perbeam.  Our favored explanation for the previous radio detection is flaring activity from a black hole low mass X-ray binary (LMXB).  We performed a new regression of the Fundamental Plane of black hole activity, valid for determining black hole mass from radio and X-ray observations of sub-Eddington black holes, finding $\log M_{\rm BH} = (1.638\pm 0.070)\log L_{\rm R} - (1.136\pm 0.077)\log L_{\rm X} - (6.863\pm 0.790)$, with an empirically-determined uncertainty of 0.44\,dex.  This constrains the mass of the X-ray source in G1, if a black hole, to be $<9.7\times10^3 M_{\odot}$ at 95\% confidence, suggesting that it is a persistent LMXB.  This annuls what was previously the most convincing evidence from radiation for an IMBH in the Local Group, though the evidence for an IMBH in G1 from velocity dispersion measurements remains unaffected by these results.
\end{abstract}


\section{Introduction}

The existence of intermediate-mass black holes (IMBHs), while theoretically well motivated \citep[e.g.][]{Vol05}, remains to be definitively confirmed.  With masses $50M_{\odot}<M_{\rm BH}<10^6 M_{\odot}$, such objects would bridge the gap between the dynamically-confirmed stellar mass black holes in the Galaxy \citep{McC06} and the supermassive black holes at the centers of galaxies \citep{Mag98}.

IMBHs are believed to form via the deaths of primordial (Population III)  stars \citep{Fry01}, or by direct collapse of large amounts of gas \citep[e.g.][]{Vol10}.  Alternatively, IMBHs can form in the cores of dense star clusters, either through successive mergers of stellar-mass black holes \citep{Mil02}, or via a runaway collision of massive stars in the cluster center followed by direct collapse \citep{Por04,Gur04}.  Thus, dense star clusters should be among the best locations to search for IMBHs, either via their effects on the dynamics of the surrounding stars, or via radiative signatures of the accretion of gas onto the black hole.

Searching for IMBHs using dynamical studies of star clusters is only possible within the Local Group \citep[e.g.][]{Set10}, where we have sufficient angular resolution to resolve the black hole sphere of influence.  Arguably the best IMBH candidate identified to date via stellar dynamics is at the center of the cluster G1 (Mayall-II) in M\,31.  With a mass of (7--17)$\times10^6M_{\odot}$ \citep{Mey01}, G1 is one of the most massive star clusters in the Local Group, and has been suggested to be the tidally stripped core of a former dwarf galaxy \citep{Ma07}.  From modeling the stellar photometry and kinematics of G1, \citet{Geb02} suggested the presence of a central IMBH of mass $2.0^{+1.4}_{-0.8}\times10^4M_{\odot}$.  This claim was challenged by \citet{Bau03}, who attributed the high central velocity dispersion to a central concentration of stellar remnants.  However, using new data, \citet{Geb05} ruled out the no-black hole model at 97\% significance, and refined the mass estimate to $1.8\pm0.5\times10^4 M_{\odot}$.

The presence of an IMBH in G1 would imply the possibility of a direct detection of its accretion signature.  Recent \xmm\ observations \citep{Tru04,Poo06} detected X-ray emission from G1 that was positionally coincident with the cluster center \citep{Kon07}, a localization subsequently confirmed to within 0.15\arcsec\ by higher-precision astrometry with \chandra\ and the {\it Hubble Space Telescope (HST)} \citep{Kon10}.  The X-ray spectrum and luminosity were consistent between the \xmm\ and \chandra\ observations, and, for a distance of 750\,kpc \citep{Vil10,Rie12} correspond to an unabsorbed 0.3--7\,keV luminosity of $2.1^{+0.7}_{-0.5}\times10^{36}$\,erg\,s$^{-1}$ \citep{Kon10}.  By considering several possible scenarios, \citet{Kon10} found that only an IMBH accreting from a companion star, or an ordinary low mass X-ray binary (LMXB) could explain the X-ray observations, although neither X-ray spectra nor X-ray astrometry could distinguish between these two cases.

From the observed correlation among radio luminosity, X-ray luminosity and mass of low-power accreting black holes \citep[the Fundamental Plane (FP) of black hole activity;][]{Mer03,Fal04}, the radio emission from an IMBH is expected to be several hundred times greater than that from an LMXB, providing an ideal discriminant between these two scenarios \citep{Mac06}.  The detection by the Very Large Array (VLA) of a $28\pm6$\,$\mu$Jy source within 1.3\arcsec\ of the center of G1 \citep[][hereafter UGH07]{Ulv07} was consistent with the radio emission expected from an accreting IMBH with the observed X-ray luminosity.  However, the relatively compact VLA configuration used meant that the astrometric accuracy was insufficient to conclusively associate the radio detection with the X-ray source.  The 1.3\arcsec\ positional offset, while formally within the 95\% confidence X-ray error circle, leaves open the possibility that the two sources are unrelated.  Furthermore, since LMXBs are known to be variable in both the radio and X-ray bands, the non-simultaneity of existing X-ray and radio data implies that the LMXB interpretation cannot be ruled out.  To address these concerns, we report here the results of deep, high-resolution, strictly-simultaneous X-ray and radio observations of G1, using \chandra\ and the Karl G. Jansky Very Large Array (JVLA).

\section{Observations and data reduction}

\subsection{Chandra}
\label{sec:xray}

{\it Chandra} observed G1 on 2011 June 26 (observation 12291) using the ACIS-S3 detector. As there was no evidence for background flaring, we analyzed the full 34587 s observation using {\sc ciao} 4.3 with {\sc caldb} 4.4.3. Using sources detected by the wavelet detection algorithm ({\sc ciao wavdetect}) and excluding G1, we registered the observation to the absolute astrometric frame used for a previous {\it Chandra} localization of G1 \citep{Kon10}. As G1 was only 0\farcm26 off-axis for our {\it Chandra} observation, we analyzed all events in a circle with a radius 50\% larger than that enclosing 90\% of the source's photons, and derived the background in a nearby annulus with 5 times the area of the source aperture.

Although a variety of spectral models were fit to the X-ray emission from G1 we only report the results for an absorbed power-law model; a power-law spectrum is expected for either an IMBH accreting at a low fraction of its Eddington luminosity or a low-luminosity LMXB. For the reported best-fit ($\chi^2=1.6$ for 3 degrees of freedom), we performed $\chi^2$ fitting after applying a minimum bin size of 15 counts and Gehrel's weighting; altering the bin size or fitting with the Cash statistic did not strongly affect the best-fit parameters. We set the absorption column density to the Galactic value of $6.5\times 10^{20} {\rm \, cm}^{-2}$ and applied the T\"ubingen-Boulder absorption (tbabs) model with abundances from \citet{Wil00} and photoelectric absorption cross-sections from \citet{Ver96}.

\subsection{JVLA}
\label{sec:jvla}

On 2011 June 26, we observed G1 for 10\,h with the JVLA \citep{Per11}, giving a total on-source time of 446\,min.  The bandwidth was split into two 1024-MHz basebands centered at 5.0 and 7.45\,GHz.  Each baseband consisted of eight 128-MHz sub-bands, each comprising 64 spectral channels of width 2\,MHz.  We observed in full polarization mode, using an integration time of 3\,s.  The JVLA was in its most-extended A-configuration, giving a maximum resolution of 0.23\arcsec.

To render our measurement insensitive to artefacts generated at the phase center by correlation errors \citep[see, e.g.][]{Eke99}, we pointed 2.5\arcsec\ south of G1.  The amplitude scale was set via observations of 3C\,48, using the coefficients derived at the JVLA by NRAO staff in 2010.  The secondary calibrator J0038+4137 was observed for one minute of every six to permit amplitude and phase calibration.  Every hour, we observed the nearby check source, J0025+3919, to estimate coherence losses via comparison of its measured flux density before and after self-calibration.

Data were reduced following standard procedures within {\sc casa} \citep{McM07}.  Imaging was carried out using natural weighting, for maximum sensitivity.  We deconvolved bright sources out to twice the distance of the half-power point of the primary beam at 5\,GHz, using the {\it w}-projection algorithm to prevent phase errors due to the sky curvature from affecting our deconvolution.  

\section{Results}

\subsection{Chandra}

With 88.8 net counts, our best fit spectrum for data in the 0.3--10 keV range had a power-law index of $\Gamma = 1.59^{+0.31}_{-0.30}$, where all error bars from X-ray spectral fitting indicate 90\% confidence intervals.  We found a centroid position for G1 of 00$^{\rm h}$32$^{\rm m}$46$^{\rm s}$.536, 39$^{\circ}$34'40''.51 (J2000), with a 1$\sigma$ positional uncertainty ellipse having semi-major and semi-minor axes of 0.18 and 0.16 arcsec, respectively. The uncertainties are dominated by systematic errors from the registration of the image to the absolute astrometric frame.

\begin{deluxetable}{llccccc}
\centering
\tabletypesize{\scriptsize}
\tablecaption{Unabsorbed fluxes and luminosities of G1 for different X-ray bands, calculated assuming a source distance of 750\,kpc. \label{tab:xbands}}
\tablewidth{0pt}
\tablehead{
\colhead{Band} & \colhead{Unabsorbed flux} & \colhead{Luminosity} \\
(keV) & ($10^{-14}$\,erg\,cm$^{-2}$\,s$^{-1}$) & ($10^{36}$\,erg\,s$^{-1}$)}
\startdata
0.5--10 & $2.59^{+0.79}_{-0.66}$ & $1.74^{+0.53}_{-0.44}$ \\
2--10 & $1.77^{+0.78}_{-0.63}$ & $1.19^{+0.53}_{-0.42}$ \\
0.3--7 & $2.30^{+0.46}_{-0.45}$ & $1.55^{+0.31}_{-0.30}$ \\
0.3--10 & $2.79^{+0.73}_{-0.63}$ & $1.88^{+0.49}_{-0.43}$ \\
\enddata
\end{deluxetable}

We measured an unabsorbed 0.5--10\,keV flux of $2.59^{+0.79}_{-0.66} \times 10^{-14} {\rm \, erg \, cm}^{-2} {\rm \, s}^{-1}$, corresponding to a luminosity of $1.74^{+0.53}_{-0.44} \times 10^{36} \, (d/750 {\rm \, kpc})^2  {\rm \, erg \, s}^{-1}$.  For comparison, we provide in Table~\ref{tab:xbands} the unabsorbed fluxes and luminosities for the different energy bands used in the literature both for G1 and to derive the FP regression coefficients.

\subsection{JVLA}

\begin{figure}
\epsscale{1.25}
\plotone{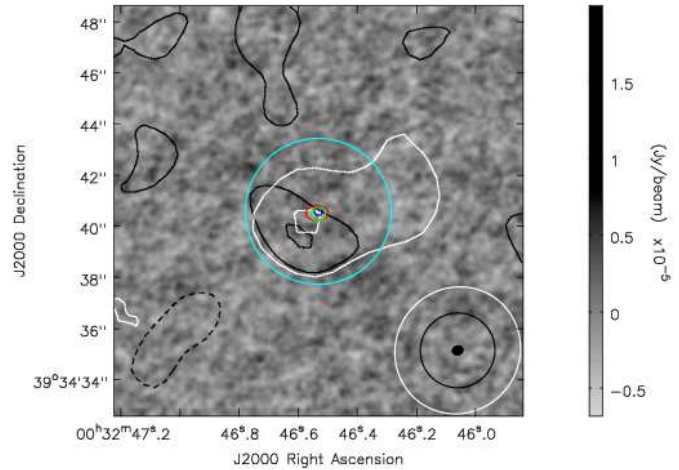}
\caption{Radio images of the field containing G1.  Greyscale shows the JVLA image from 2011 June 26.  Contours show our images of the archival VLA data (black contours for 2006 November 24--26, white contours for 2007 January 14--15), with contours at intervals of $\pm2,4$ times the rms noise (7.2 and 16.3\,$\mu$Jy\perbeam, respectively).  Blue and cyan error circles denote the core and half-mass radius of the cluster, respectively (position from the {\it HST} data of \citet{Kon10} and radii from \citet{Ma07}), green circle denotes the 95\% confidence error circle of \citet{Kon10}, and red ellipse denotes the 95\% confidence error ellipse from our \chandra\ observations.  Beam sizes for the radio images are shown at lower right (filled circle for the JVLA data, and open circles for the VLA data).  The JVLA observations detected no radio emission within 28\arcsec\ of the cluster core.\label{fig:evla_image}}
\end{figure}

No radio source was significantly detected within 28\arcsec\ of the center of G1.  Our final image (Figure~\ref{fig:evla_image}), made using the data from both basebands, had a noise level of 1.58\,$\mu$Jy\perbeam, within 9\% of the theoretically-predicted thermal noise.  The check source showed that coherence losses were $<1.5$\%.  Thus, our $3\sigma$ upper limit on the radio luminosity of G1 on 2011 June 26, assuming a flat radio spectrum extending up to the central frequency of 6.225\,GHz, is $<2.0\times10^{31}$\,erg\,s$^{-1}$.

Given the results of UGH07, we can propose three possible explanations for our highly-constraining radio non-detection.  First, the emission could be extended (e.g.\ a pulsar wind nebula) and therefore resolved out in the higher-resolution JVLA observations.  Alternatively, the radio source could be variable, by a factor of $\gtrsim6$, on a timescale of years.  Finally, the detection reported by UGH07 could have been an artefact, arising either from the mix of VLA and JVLA antennas in use in 2006, or from the unfortunate coincidence of a noise spike with the known X-ray position.

To test the first scenario, we tapered our JVLA data to the resolution of the C-configuration data of UGH07.  No significant emission was detected within the 95\% radio error circle, down to an rms noise level of $5.6$\,$\mu$Jy\perbeam, thereby ruling out extended emission.

\subsection{Reanalysis of archival VLA data}

\begin{deluxetable*}{llcccccc}
\centering
\tabletypesize{\scriptsize}
\tablecaption{Radio observations of G1.  Where two different flux densities are quoted, the first is for the full data set, whereas the second excludes the newly-upgraded JVLA antennas. \label{tab:vla}}
\tablewidth{0pt}
\tablehead{
\colhead{Program code} & \colhead{Date} & \colhead{Array config.} & \colhead{Frequency} & \colhead{Bandwidth} & \colhead{Time on source} & \colhead{Source flux density} & \colhead{Image noise} \\
& & & (GHz) & (MHz) & (min) & ($\mu$Jy\perbeam) & ($\mu$Jy\perbeam)}
\startdata
AA 276 & 2002 Sep 11 & B & 8.4 & 100 & 85 & & 47\\
AG 730 & 2006 Nov 24--26 & C & 8.4 & 100 & 845 & 15/31 & 5.5/7.2\\
AK 634 & 2007 Jan 13--14 & C & 4.9 & 100 & 214 & & 16/22\\
AU 116 & 2007 Jan 14--15 & C & 4.9 & 100 & 235 & 67/63 & 16/20\\
SC 489 & 2011 Jun 26 & A & 6.2 & 2048 & 446 & & 1.6\\
\enddata
\end{deluxetable*}

To investigate radio variability, we reanalyzed the archival VLA observations of G1 (Table~\ref{tab:vla}).  In the observations from 2006 and 2007, 5--6 of the original VLA antennas had been retrofitted with the new JVLA electronics.  To account for the differing bandpasses of the retrofitted JVLA antennas, we applied baseline corrections within {\sc aips} \citep{Gre03}, after which we imported the data into {\sc casa} for further calibration and imaging, using the same procedures and calibrator sources as for the JVLA data (Section~\ref{sec:jvla}).  UGH07 noted some issues with the JVLA antennas in the 8.4-GHz observations of 2006 November, which we found to be manifested as large (albeit smoothly-varying) phase corrections.  As a conservative approach, we therefore made images both including and excluding data from the JVLA antennas from all observations from 2006 and 2007.  At 4.9 (but not at 8.4) GHz, there was sufficient emission in the field to apply phase self-calibration with a 10-minute solution interval.

Only two data sets showed marginally-significant radio emission at the originally-reported position (Table~\ref{tab:vla}).  In the 8.4-GHz observations of 2006 November, when excluding the JVLA antennas we found a $4.3\sigma$ source at $31\pm7$\,$\mu$Jy\perbeam, consistent with the results of UGH07.  However, enhanced local noise in the source region (rms of 8.3\,$\mu$Jy\perbeam) reduces the true significance to $3.8\sigma$.  The detection was not significant when the JVLA antennas were included.

In the 4.9-GHz observations of 2007 January 14--15, there were no noticeable issues with the JVLA antennas.  A $4.1\sigma$ source was detected at a position consistent with the radio source reported by UGH07 ($3.2\sigma$ when excluding the JVLA antennas), although again the local rms appeared to be slightly enhanced.  The data taken the previous day (January 13--14) showed no significant emission at the same position.  Thus, either the radio emission is variable on timescales of 1 day, or the marginally-detected source is spurious.

\section{Discussion}

\subsection{Variability}
\label{sec:variability}

While our archival re-analysis confirms the measurement of UGH07, the enhanced local noise close to the source position reduces its significance.  Coupled with the phase-center location and the slight positional offset from the well-determined X-ray position \citep{Kon10}, we cannot exclude the possibility that it is a noise spike.  However, the probability of $4\sigma$ noise spikes coinciding in two independent images (Fig.~\ref{fig:evla_image}) makes this interpretation unlikely.

Should these two marginal detections be real, they would imply a variation in the radio emission by a factor of $\gtrsim6$ between 2006 and 2011, and variations on timescales as short as 1\,day in 2007 January.  Radio emission at this level ($\sim10^{32}$\,erg\,s$^{-1}$) could correspond to sub-Eddington flaring from an IMBH, or to a major outburst of a black hole LMXB (an outbursting neutron star would not reach the observed radio luminosity).  Stellar-mass black holes are known to exhibit short-timescale, sub-Eddington radio flares \citep[e.g.][]{Mil08}.  Should the radio emission be an analogous flare from an IMBH, then scaling the variability timescale with accretor mass would imply radio variations over weeks to months rather than days, making this scenario unlikely.

The more likely explanation is an LMXB outburst.  Given the observed correlation between outburst duration, peak luminosity, and orbital period in LMXBs \citep{Por04b}, the radio luminosity would imply an extended period (several months) above $10^{36}$\,erg\,s$^{-1}$.  Bright radio flares occurring over a period of a few months and varying on timescales of days have been seen in black hole LMXBs \citep[e.g.][]{McC09}.  The lack of X-ray coverage between 2003 and 2008 implies that the corresponding X-ray flare would not have been detected.

\subsection{Mass constraints from the Fundamental Plane}

Although black hole masses can in principle be constrained using the FP relationship \citep{Mer03}, the error bars on the best-fitting regression coefficients are relatively large. Furthermore, the regression is sample-dependent, with the intrinsic scatter about the FP being minimized when using only strongly sub-Eddington objects \citep{Koe06}.  While most published FP relations \citep{Mer03,Koe06,Plo12} have used radio or X-ray luminosity as the dependent variable, the intrinsic scatter about the plane implies that it is not valid to simply invert the best-fitting regression coefficients to determine a mass constraint from the measured radio and X-ray luminosities \citep{Plo12}.  The sole existing mass regression \citep{Gul09} included only supermassive black holes ($M>10^6M_{\odot}$), so has limited predictive power for lower-mass systems. 

\begin{figure}
\epsscale{1.1}
\plotone{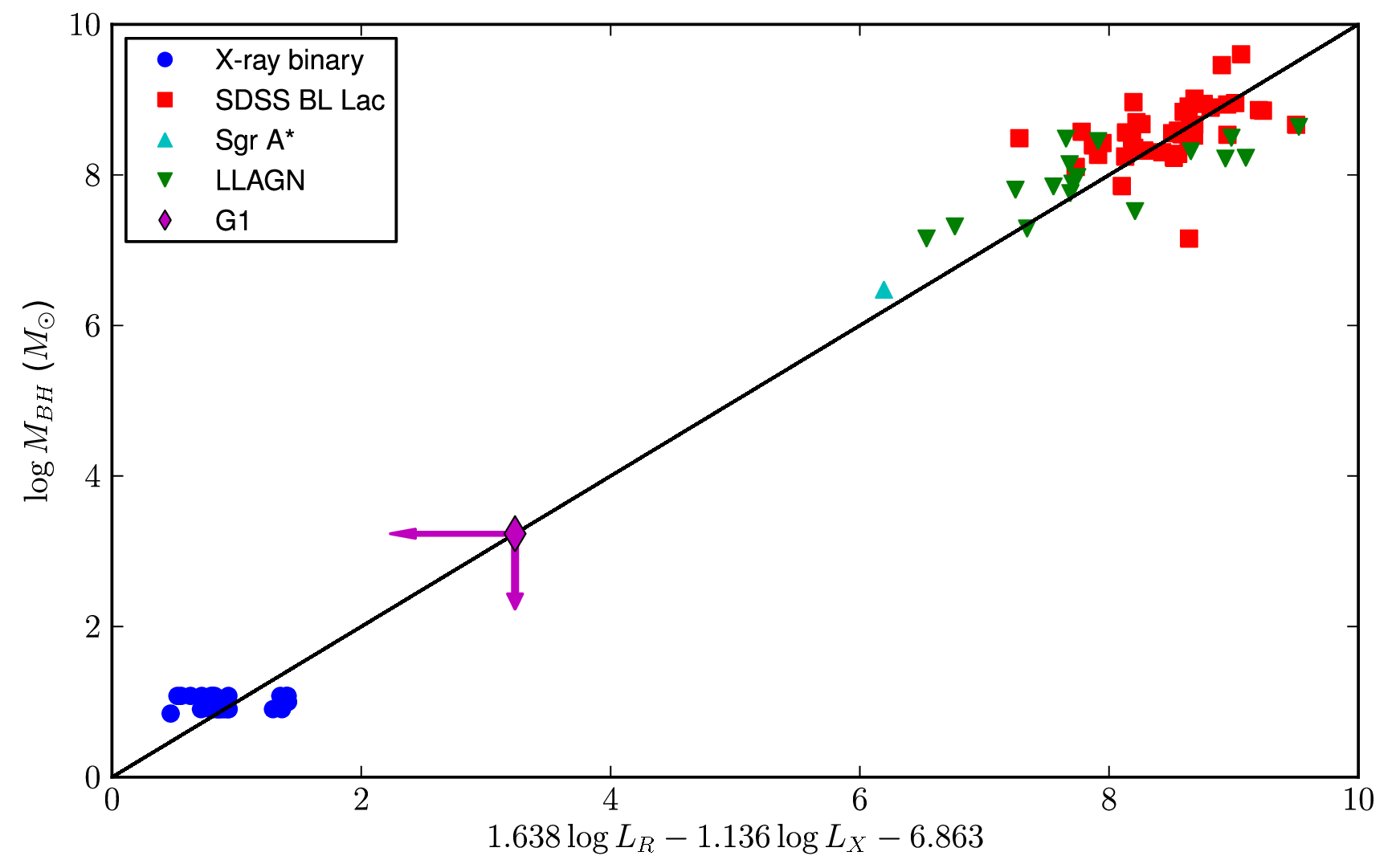}
\caption{Our best-fitting regression for the FP (Equation~\ref{eq:fp}) as a function of black hole mass, showing the limits implied by our simultaneous radio and X-ray observations of G1.\label{fig:fplane}}
\end{figure}

\begin{figure}
\epsscale{1.1}
\plotone{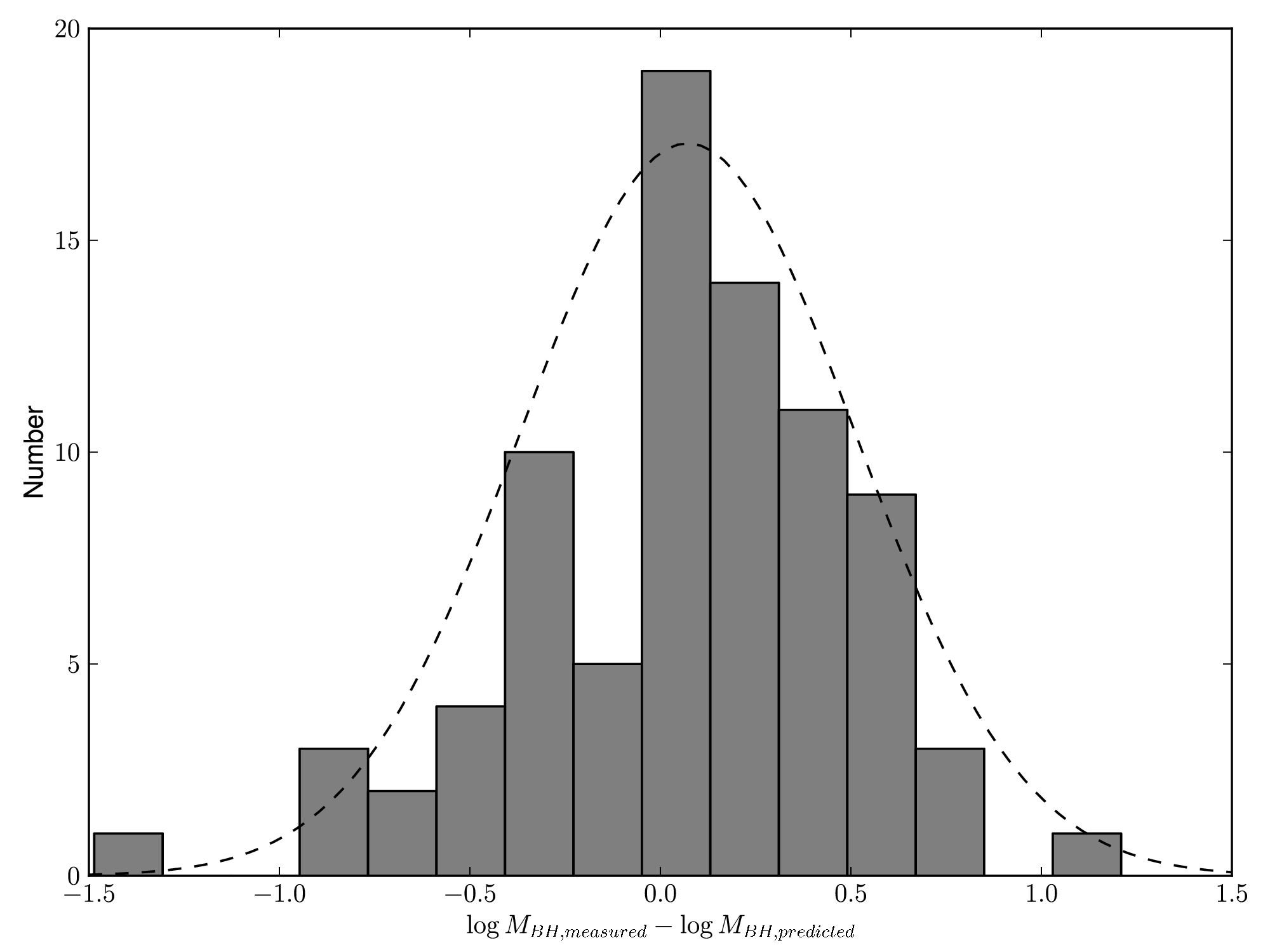}
\caption{Histogram of the residuals from our best-fitting FP relation (Equation~\ref{eq:fp}), together with a Gaussian of the same mean and standard deviation (0.44\,dex).  The scatter about the best-fitting regression implies that black hole masses cannot be predicted to better than a factor of $\sim3$ using the FP, and that FP mass predictions can be in error by over an order of magnitude.\label{fig:fp_resid}}
\end{figure}

To place a valid constraint on the mass of the X-ray source in G1, we therefore performed a new regression of the sub-Eddington sample of \citet{Plo12}, with mass as the dependent variable.  The inclusion of BL Lacs in the sample (having controlled for synchrotron cooling effects) makes the implicit assumptions that the radio emission has a flat or inverted spectrum, and that the X-rays are predominantly optically-thin synchrotron emission.  However, it improves the dynamic range in black hole mass, and any systematics introduced by relativistic beaming are of secondary importance when compared with the intrinsic scatter about the FP. Following \citet{Plo12}, we used the Bayesian technique of \citet{Kel07} to handle the coupled uncertainties in radio and X-ray luminosities.  We find a best-fitting regression
\begin{equation}
\begin{split}
\log M_{\rm BH} &= (1.638\pm 0.070)\log L_{\rm R} \\
&- (1.136\pm 0.077)\log L_{\rm X} - (6.863\pm 0.790),
\end{split}
\label{eq:fp}
\end{equation}
where $L_{\rm R}$ and $L_{\rm X}$ are the radio and 0.5--10\,keV X-ray luminosities, respectively, in erg\,s$^{-1}$.  Our simultaneous measurements then predict a black hole mass of $<1.7\times10^3M_{\odot}$ in G1 (Figure~\ref{fig:fplane}).  Comparing measured black hole masses with those predicted by the regression shows that the $1\sigma$ uncertainty on the mass estimates is 0.44\,dex (Figure~\ref{fig:fp_resid}).  This is significantly higher than the 0.12\,dex used by previous authors to determine the uncertainties on FP-determined IMBH masses \citep[e.g.][]{Cse10,Kir12}, but, being empirically-determined from the data, should be a more accurate estimate of the true uncertainty.  Thus, we estimate the 95\% confidence upper limit on the mass of the X-ray source in G1, if a black hole, to be $<9.7\times10^3M_{\odot}$.

\subsection{Does the X-ray emission arise from an IMBH?}

\citet{Kon10} concluded that the only possible explanations for the X-ray emission from G1 were an IMBH accreting from a companion star, or a typical LMXB.  As discussed by \citet{Poo06}, seven persistent LMXBs are known in Galactic globular clusters \citep{Ver06} at X-ray luminosities $\gtrsim10^{36}$\,erg\,s$^{-1}$.  Since G1 is a particularly massive, dense cluster \citep{Mey01}, it likely contains many LMXBs. The stability of the X-ray luminosity between 2001 and 2011, coupled with the positional coincidence between the \chandra\ detections of 2008 and 2011, makes it unlikely (albeit not impossible) that some fraction of the X-ray emission arises from multiple LMXBs in the cluster core.  However, should LMXBs contribute some or all of the observed X-ray luminosity, then our upper limit on the mass of an IMBH in G1 would increase, via Equation~\ref{eq:fp}.

Given the observed X-ray count rate, there is no way to distinguish the X-ray spectrum of an IMBH from an LMXB with current instruments.  Combining the lack of strong radio emission from G1 with the fact that LMXBs of the observed luminosity and spectral shape are relatively common in globular clusters like G1, an LMXB origin for the X-ray emission is the most probable interpretation.  However, given the positional offset from the marginal radio detections of 2006--2007, the radio and X-ray emission likely arise from different LMXBs.

\subsection{Intermediate-mass black holes in globular clusters}

Although IMBHs are predicted to exist in globular clusters \citep[e.g][]{Mil02}, many attempts to detect central dark masses via stellar kinematics \citep[e.g.][]{Ger03,Noy08} have been controversial \citep[see][respectively]{Bau03a,vdM10}.  While deep radio observations have the potential to discriminate between an IMBH and a collection of dark remnants \citep{Mac04}, our result implies that no radio observations to date have shown conclusive evidence for an IMBH in a globular cluster \citep{Mac08}.  The most stringent radio limits available imply that either IMBHs are rare in globular clusters, or that they are extremely inefficient accretors \citep{Str12}.

\section{Conclusions}

Deep radio continuum observations with the JVLA detected no radio emission within 28\arcsec\ of the X-ray source at the center of G1, to a $3\sigma$ level of 4.7\,$\mu$Jy\perbeam.  The 0.5--10\,keV X-ray luminosity measured simultaneously with \chandra\ was $1.74^{+0.53}_{-0.44} \times 10^{36} \, (d/750 {\rm \, kpc})^2  {\rm \, erg \, s}^{-1}$.  Using these measurements together with a new FP regression, we constrain the mass of the X-ray source in G1, if a black hole, to be $<9.7\times10^3M_{\odot}$, at 95\% confidence.  Our $3\sigma$ radio upper limit is a factor of 6 deeper than the previously-reported VLA detection (UGH07), suggesting either that the previously-detected source was an artefact, or, more likely, that the radio emission is time-variable, and arises from a black hole LMXB in outburst.

While the FP correlation works well for a large sample of sources, it is a statistical tool, and as Fig.~\ref{fig:fp_resid} shows, can be in error by over an order of magnitude for individual sources.  Therefore, despite our radio non-detection, we cannot definitively exclude the X-ray source in G1 being an IMBH. Thus, the $4\sigma$ dynamical detection \citep{Geb05} still makes it a viable globular cluster IMBH candidate.  Our current radio limits are unlikely to be surpassed in any reasonable integration time with the current generation of telescopes, so unless the possible variability can be confirmed, a definitive identification of the X-ray source will remain the preserve of future instruments.

\acknowledgments The National Radio Astronomy Observatory is a facility of the National Science Foundation operated under cooperative agreement by Associated Universities, Inc.  GRS and COH are supported by NSERC, and COH also by an Ingenuity New Faculty Award. RMP acknowledges support from an NWO Vidi Fellowship.  This research has made use of NASA's Astrophysics Data System.

\end{document}